\documentclass[11pt]{iopart}
\usepackage{graphicx}
\begin{document}

\title[Temperature-induced emergence of Wigner correlations\ldots]{Temperature-induced emergence of Wigner correlations in a STM-probed one-dimensional quantum dot}

\author{N Traverso Ziani,$^{1,2}$ F Cavaliere$^{1,2}$ and M Sassetti$^{1,2}$}
\address{
$^1$ Dipartimento di Fisica, Universit\`a di Genova, Via Dodecaneso 33,
 16146, Genova, Italy.\\
\noindent $^2$ CNR-SPIN, Via Dodecaneso 33,
 16146, Genova, Italy.
}
\ead{fabio.cavaliere@gmail.com}

\begin{abstract}
The temperature-induced emergence of Wigner correlations over finite-size effects in a strongly interacting one-dimensional quantum dot are studied in the framework of the spin coherent Luttinger liquid. We demonstrate that, for temperatures comparable with the zero mode spin excitations, Friedel oscillations are suppressed by the thermal fluctuations of higher spin modes. On the other hand, the Wigner oscillations, sensitive to the charge mode only, are stable and become more visible. This behavior is proved to be robust both in the thermal electron density and in  the linear conductance in the presence of an STM tip. This latter probe is not directly proportional to the electron density and may confirm the above phenomena with complementary and additional information.
\end{abstract}

%\pacs{{73.21.La}, {71.10.Pm}, {73.63.-b}, {73.22.Lp}}
\maketitle

\section{Introduction}
\label{sec:intro}
Wigner crystallization~\cite{wigner} is among the most striking quantum many body effects. A Wigner crystal can emerge in two and three dimensions when the interaction energy among electrons dominates over the kinetic one~\cite{vignale}. In one dimension (1D), due to quantum fluctuations, a true crystal cannot be established, but when the correlation length exceeds the size of the sample, electrons can form a so called Wigner molecule, the finite counterpart of the Wigner crystal. The formation of Wigner molecules is not limited to the 1D case, but has been intensively studied in two dimensions as well~\cite{wigmol1,wigmol2}.\\
An ideal playground for studying correlated electrons, including Wigner molecules, are quantum dots~\cite{koudots}. Electronic correlations have been extensively studied in two dimensional quantum dots, mainly using numerical techniques,~\cite{wigmol1,wigmol2,2Dnum2,Egger99,2Dnum3,2Dnum5,2Dnum6,2Dnum7,serra,2Dnum8,2Dnum10} and the emergence of the Wigner molecule has been demonstrated. Due to the high level of symmetry of the typical circular two dimensional quantum dots, the density profile of the Wigner molecule is rotationally invariant. A major implication of this issue is that the electron density is not effective in the characterization of the Wigner molecule~\cite{wigmol1,wigmol2,2Dnum8} and attempts at imaging the correlated electron wavefunction employing a scanning tunnel microscope (STM) have been put forward~\cite{maxwf}.

\noindent In finite size 1D systems the scenario is different~\cite{giamarchi,schulz,kramer,xia,bortz,1Dwig}. Here, due to the broken translational invariance, the formation of the Wigner molecule can be investigated via the electron density. Indeed it has been shown that increasing interaction or decreasing the density, a crossover in the density between the Friedel and Wigner oscillations occurs. This result is supported both by analytical~\cite{giamarchi,schulz,bortz,mantelli,safi,sablikov} and numerical calculations~\cite{kramer,xia,bortz,pederiva,bedu,szafran,wire3,polini,shulenburger,secchi1,sgm2,astrak,burke,polini2,silva}.

\noindent Recently, in order to study the density oscillations arising due to the formation of a Wigner molecule, the properties of 1D quantum dots coupled to AFM tips have been explored,~\cite{sgm2,mantelli,AFM,sgm1,linear} a set up proposed~\cite{lee,glazcap} also for the detection of spin charge separation,~\cite{giamarchi,sassetti98}
another hallmark of one-dimensional electron systems.
Coupling quantum dots to STM tips has also been proposed. It was indeed suggested in order to
detect vibrations~\cite{noi} and  Friedel oscillations ~\cite{eggertstm,martin,dolcini,nocera} and more recently for the detection of the Wigner molecules~\cite{secchi2}.
As long as this last issue is concerned, the investigation relied on exact diagonalization performed numerically, and hence restricted to the case of few electrons only.

\noindent In order to extend the numerical results to many particles, analytical models must be employed. The Luttinger liquid theory~\cite{giamarchi,haldanefluid,voit,delft}
has been successfully adopted in the description of 1D quantum dots,~\cite{eggerimp,braggioepl,iotobias,kim,ioale1,milena} even in the Wigner crystal regime~\cite{schulz,sablikov,safi,bortz}. It represents the low energy sector of 1D microscopic models, ranging from the Hubbard one~\cite{bortz,k05} to ones directly related to the Wigner molecule, in which electrons oscillate around their equilibrium positions~\cite{fiete1,075}.

\noindent In this work we study the effects of temperature on the Wigner and Friedel correlations in a 1D quantum dot. The temperature range of our investigation is such that $k_BT\ll D_\sigma \ll D_\rho$, where $D_{\rho/\sigma}$ are the charge and spin bandwidth, and hence the dot is  modeled within the coherent Luttinger liquid picture~\cite{giamarchi,fiete1}. We demonstrate that Wigner correlations emerge over the Friedel ones when temperature is raised up to the temperature of the spin addition energy $E_\sigma$~\cite{pscr}. The reason for this emergence is the suppression of the Friedel oscillations: when spin excitations are thermally activated Friedel oscillations with different wavelength superimpose and result in an overall suppression of their amplitude. On the other hand, Wigner oscillations are robust against the increase of temperature, since they are not sensitive to the spin excitations but to the charge ones which lie, in presence of strong interaction, at a much higher excitation energy.
We prove this behavior by studying the thermal electron density and the linear conductance in the presence of an STM tip. This latter probe is chosen since it is not directly proportional to the electron density and supports the universality of the above phenomena, providing independent information.

\noindent The scheme of the paper is the following. In Sec.~\ref{sec:model} we introduce the model of the quantum dot within the Luttinger framework and we evaluate the temperature dependence of the equilibrium electron density. In Sec.~\ref{sec:transport} we evaluate the linear conductance in the presence of an external STM tip. The last part of the section contains the main results, that are focused on the stability of Friedel and Wigner oscillations as a function of temperature. Finally, in Appendix~\ref{sec:appa}, we outline the calculation of the tunneling rates.
\section{Model}
\label{sec:model}
\subsection{Isolated quantum dot}
We consider a 1D quantum dot of length $L$, treated within the coherent Luttinger liquid description~\cite{giamarchi,voit} with linearized bands around the Fermi points.
The dot Hamiltonian $H_{d}$ reads as ($\hbar=1$)
\begin{equation}
H_{d}=H_\rho+H_\sigma+H_{b},\label{eq:toth}
\end{equation}
with
\begin{eqnarray}
H_\rho&=&\frac{E_{\rho}}{2}( N_{\rho}-N_\rho^{(0)}-N_g)^2, \\
H_\sigma&=&\frac{E_{\sigma}}{2}N_{\sigma}^2,\\
H_{b}&=&\sum_{n_{q}>0}\left[\varepsilon_{\rho}n_{q}d^\dag_{\rho,n_{q}}d_{\rho,n_{q}}+\varepsilon_{\sigma}n_{q}d^\dag_{\sigma,n_{q}}d_{\sigma,n_{q}}\right]\, . \label{eq:hp}
\end{eqnarray}
Here, $E_\rho$ and $E_\sigma$ represent the charge and spin addition energies of the zero modes with
\begin{equation}
N_{\rho/\sigma}= N_{s=+}\pm N_{s=-}\, ,
\end{equation}
and $N_{s}$ the total electron number with spin $s$. $N^{(0)}_{\rho}$ represents the reference number of electrons, chosen to be even
with an equal number of electrons per spin direction $N_+^{(0)}=N_-^{(0)}$.

\noindent The addition energies $E_{\nu}=\pi v_{\nu}/2Lg_{\nu}^2$ ($\nu\in \{\rho,\sigma\}$) are proportional to the propagation velocities $v_{\nu}$ of the modes and scale with the interaction parameters $g_{\nu}$. For repulsive interactions one has $g_{\rho}=g<1$, while $g=1$ corresponds to the noninteracting
limit. On the other hand, $g_{\sigma}=1$ for an SU(2) invariant
theory~\cite{voit}. Furthermore, $N_{g}$ is the number of charges
induced by the gate capacitively coupled to the dot.
The term $H_{b}$ describes collective, quantized charge and
spin density waves with boson operators $d_{\nu,n_q}$, with $n_{q}$ a non-negative integer, and
$\varepsilon_{\nu}=\pi v_{\nu}/L$. As long as the velocities are concerned, one has for strongly interacting electrons  $v_\sigma\ll v_\rho$ and hence $2E_\sigma=\varepsilon_{\sigma}\ll \varepsilon_{\rho,}E_\rho$, while in the noninteracting regime $v_\rho=v_\sigma=v_F$, with $v_F$  the Fermi velocity. The wide separation or $v_{\sigma}$ and $v_{\rho}$ occurring in the regime of strong interactions is the ultimate responsible for the effects we are going to discuss.\\
We remind that the coherent Luttinger liquid approach describes the low energy sector of interacting electrons as long as the external parameters, such as voltages and/or temperatures, are low: $|eV|,k_BT\ll D_\sigma,D_\rho$, with $D_{\rho/\sigma}=N_\rho E_{\rho/\sigma}$ the band width of the spin and charge sectors respectively. The regime of higher temperatures or voltages can be described within the so called incoherent Luttinger liquid model \cite{fiete1,075} which however will not be not treated in this work.\\

The field operator $\psi_s(x)$ with spin $s$, satisfying open boundaries conditions $\psi_s(0)=\psi_s(L)=0$ is
given by~\cite{fabrizio}
\begin{equation}
\psi_s(x)=e^{ik_Fx}\psi_{s,+}(x)+e^{-ik_Fx}\psi_{s,-}(x)\, , \label{eq:optot}
\end{equation}
with Fermi wave vector $k_F=N_s^{(0)}\pi/L$ and $2L$-periodic fermion fields $\psi_{s,r}(x)$ with $r=\pm$ representing right and left movers. Due to the open boundary conditions one has
$\psi_{s,r}(x)=-\psi_{s,-r}(-x)$, with the bosonic representation~\cite{fabrizio}
\begin{equation}
\psi_{s,+}(x)=\frac{\eta_s}{\sqrt{2\pi\alpha}}e^{-i\theta_{s}}
\,e^{i\frac{\pi \Delta N_sx}{L}}e^{i\frac{\Phi_{\rho}(x)+s\Phi_\sigma(x)}{\sqrt{2}}}\,. \label{eq:opright}
\end{equation}
Here, $\Delta N_s=N_s-N_s^{(0)}$, and $\alpha$ is the cutoff length assumed to be the inverse of the Fermi momentum ($\alpha=k_F^{-1}$). The operator $\theta_{s}$ satisfies $[\theta_{s}, N_{s'}]=i\delta_{s,s'}$, and the relation $\eta_s\eta_{s'}+\eta_{s'}\eta_s=2\delta_{s,s'}$ allows the correct
anticommutation relations among the fields with different spin. The
boson fields $\Phi_{\rho}(x)$, $\Phi_{\sigma}(x)$ are given by ($q=n_q\pi/L$)
\begin{equation}
\Phi_{\nu}(x)\!=\!\sum_{n_{q}>0}\frac{e^{-\alpha q/2}}{\sqrt{g_{\nu}n_q}}
\left\{\left[\cos{(qx)}-ig_{\nu}\sin{(qx)}\right]d^\dag_{\nu,n_{q}}+\mathrm{h.c.}\right\}\,\nonumber
\end{equation}
\subsection{Electron density}
In this subsection we would like to focus on the behavior of the total electron density $\rho(x)=\sum_s \rho_s(x)$.
As already pointed out~\cite{safi,schulz,haldanefluid,mantelli} the density consists of a series of contributions that can be expressed via bosonization. Here we will take into account the most relevant ones
\begin{equation}
\rho(x)=\rho^{LW}(x)+(1-\lambda)\rho^F(x)+\lambda\rho^W(x). \label{eq:den2}
\end{equation}
The first term represents the long wave contribution $\rho^{LW}(x)=\sum_{r,s}\psi^{\dagger}_{s,r}(x)\psi_{s,r}(x)$
with bosonized form
\begin{equation}
\rho^{LW}(x)=\frac{2k_F}{\pi}+\frac{\Delta N_{\rho}}{L}-\frac{\sqrt{2}}{\pi}\partial_x \varphi_\rho(x)\,,
\end {equation}
where
 $\Delta N_\rho=\Delta N_{s=+}+\Delta N_{s=-}$ and
\begin{equation}
\varphi_{\rho/\sigma}(x)=\frac{1}{2}\left[
 \Phi_{\rho/\sigma}(-x)-\Phi_{\rho/\sigma}(x)\right].\label{eq:varphi} \end{equation}
The Friedel term $\rho^F(x)=\sum_{s}\left(e^{-2ik_{F}x}\psi^{\dagger}_{s,+}(x)\psi_{s,-}(x)+ \mathrm{h.c.}\right)$,
induced by the finite size of the dot, is responsible for $2k_F$ oscillations~\cite{fabrizio}\footnote{Other forms of the density operator conserving the total number of particles~\cite{haldanefluid,sablikov,mantelli} lead to results that only
slightly differs at the dot boundaries from the one we use.}
\begin{eqnarray}
\rho^F(x)&=&\sum_s\rho_s^F(x),\\
\rho_s^F(x)&=&-\frac{ N_s}{ L}\cos\left[\mathcal{L}(\Delta N_{s},x)-2\varphi_{s}(x)\right]\, ,\\
\mathcal{L}(n,x)&=&2k_{F} x+\frac{2\pi x}{L}n -2h(x)\, ,
\end{eqnarray}
with
\begin{eqnarray}
\varphi_s(x)&=&\frac{\varphi_\rho(x)+s\varphi_\sigma(x)}{\sqrt{2}},\\
h(x)&=&\frac{1}{2}\tan^{-1}\left[\frac{\sin(2\pi x/L)}{e^{\pi\alpha/L}-\cos (2\pi x/L)}\right]\, .\label{eq:effe}
\end{eqnarray}

The last term in Eq.~(\ref{eq:den2}) is the Wigner contribution
\begin{equation}
\rho^W\!\!(x)\!\!=\!\!\pi\alpha e^{-4ik_Fx}\psi^\dag_{+,+}\!(x)\psi_{+,-}\!(x)\psi^\dag_{-,+}\!(x)\psi_{-,-}\!(x)+\mathrm{h.c.},
\end{equation}
which may arise due to interaction effects, band curvature or other external perturbations~\cite{sablikov,safi}.
It is responsible for the $4k_F$ oscillations and has the form
\begin{equation}
\rho^W\!\!(x)=-\frac{N_\rho}{L}\cos\!\left[2\mathcal{L}(\Delta N_{\rho}/2,x)-2\sqrt{2}\varphi_\rho(x)\right]\, .
\end{equation}
Note that, in contrast to the Friedel term, the Wigner one depends on the charge sector only. The parameter $\lambda\in[0,1]$ ensures the right boundary values of the density operator ($\rho(0)=\rho(L)=0$), and depends both on the interaction strength and on the average electron density. In particular, for strong interactions and/or low densities one expects $\lambda\to 1$, while for weakly interacting systems and/or high densities $\lambda\approx 0$ as numerical studies suggest~\cite{bortz}.\\

\begin{figure}[htbp]
\begin{center}
\includegraphics[width=12cm,keepaspectratio]{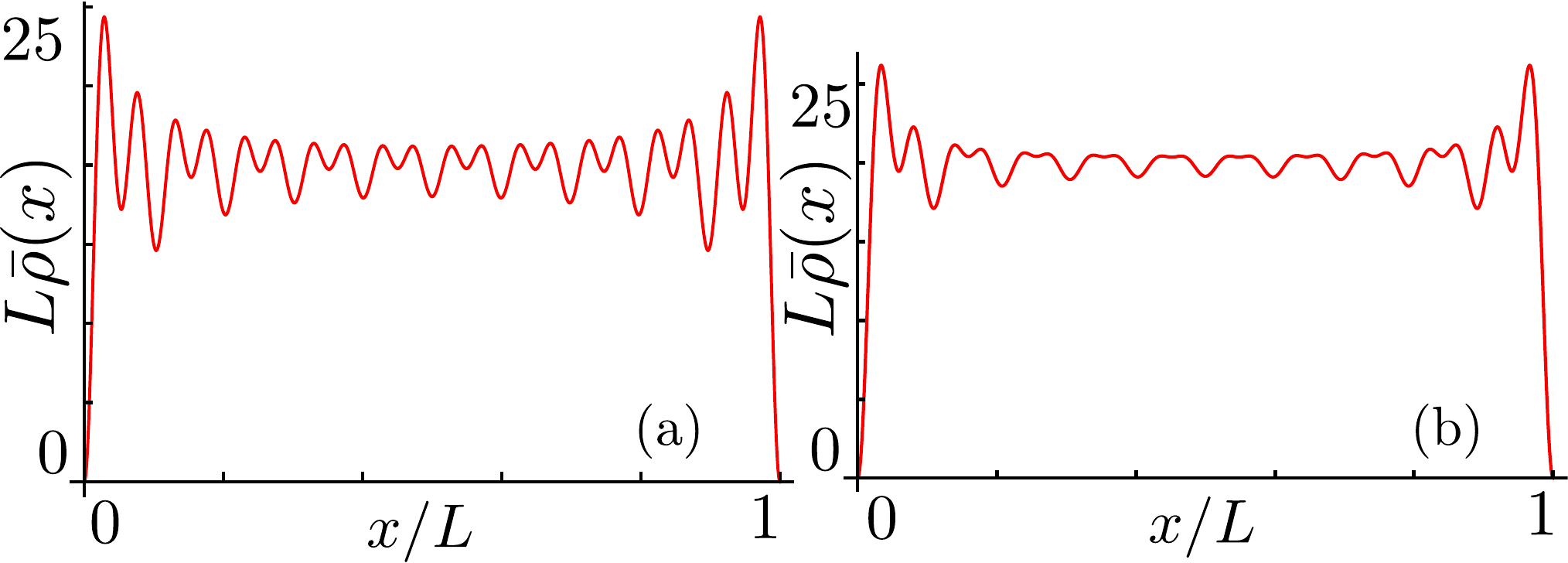}
\caption{(Color online) Adimensionalized electron density as a function of the position $x/L$ with $N^{(0)}_{\rho}=20$, $T=0$ and for
 (a) $g=0.5$; (b) $g=0.8$. In all panels $v_\rho=40v_\sigma$ and $\lambda=0.7$.}
\label{fig:densita1}
\end{center}
\end{figure}

Let us now evaluate the thermal average of the electron density. The quantum dot is assumed in thermal equilibrium at fixed total electron's number $N_\rho$. The corresponding excited zero spin modes $N_{\sigma}$ range from $-N_\rho\le N_\sigma\le N_\rho$. They are weighted by the thermal probability $e^{-\beta H_\sigma}/Z_{\sigma}$. Charge and spin density waves are also distributed with probability $e^{-\beta H_{b}}/Z_b$ ($Z_{b/\sigma}$ the partition functions associated to $H_{b/\sigma}$).
The thermal equilibrium density $\bar{\rho}(x)$ can then be written as
\begin{equation} \!\!\!\bar{\rho}(x)=\!\!\!\!\sum_{N_\sigma=-N_\rho}^{N_\rho}\!\!\!\!\frac{\langle N_{\rho},N_{\sigma}| Tr_{b}  e^{-\beta H_{b}}e^{-\beta H_\sigma} \rho(x) |N_{\rho},N_{\sigma}\rangle}{Z_{b}Z_{\sigma}},
\end{equation}
where the trace is over the bosonic charge and spin degrees of freedom. Inserting Eq.~(\ref{eq:den2}) in the above expression one obtains the following contributions
\begin{equation}
\bar{\rho}(x)=\frac{2k_F}{\pi}+\frac{\Delta N_\rho}{L}+(1-\lambda)\sum_s\bar{\rho}^F_s(x)+\lambda\bar{\rho}^W(x),
\label{densita1}
\end{equation}
with
\begin{eqnarray}
\bar{\rho}^F_s(x)&=&\!\!\!\!\!\!\sum_{N_\sigma=-N_\rho}^{N_\rho}\!\!\!\!\frac{\langle N_{\sigma}|(N_\rho-s N_\sigma)e^{-\beta H_\sigma}|N_{\sigma}\rangle}{2Z_\sigma}\bar{\rho}_s(x),\label{densita2}\\
\!\!\bar{\rho}_s(x)\!&=&\!-\cos\left[\mathcal{L}\left(\frac{\Delta N_\rho-s N_\sigma}{2},x\right)\right]\!\!\frac{e^{-2\langle\varphi_s^2(x)\rangle}}{L},\\
\bar{\rho}^{W}(x)&=&-\frac{N_\rho}{L}\cos\left[2\mathcal{L}\left(\frac{\Delta N_\rho}{2},x\right)\right]e^{-4\langle\varphi_\rho^2(x)\rangle}, \\
\!\!\!\langle\varphi_{s/\rho}^2(x)\rangle &=& T r_b\frac{e^{-\beta H_b}}{Z_b}\varphi_{s/\rho}^2(x).\label{densita4}
\end{eqnarray}
Let us now recall the behavior at $T=0$, already discussed~\cite{mantelli}. At zero temperature the allowed spin numbers in the sum of Eq.~(\ref{densita2}) are $N_{\sigma}=0$ or $N_{\sigma}=\pm 1$ for even/odd electrons. In addition, the trace in Eq.~(\ref{densita4}) can be evaluated giving
\begin{eqnarray}
e^{-2\langle\varphi_s^2(x)\rangle}&=&\left[\frac{\sinh\left(\frac{\pi\alpha}{L}\right)}{\sqrt{\sinh^2\left(\frac{\pi\alpha}{L}\right)+\sin^2\left(\frac{\pi x}{L}\right)}}\right]^{\frac{1+g}{2}},\label{densita5a}\\
e^{-4\langle\varphi_\rho^2(x)\rangle}&=&\left[\frac{\sinh\left(\frac{\pi\alpha}{L}\right)}{\sqrt{\sinh^2\left(\frac{\pi\alpha}{L}\right)+\sin^2\left(\frac{\pi x}{L}\right)}}\right]^{2g}.
\label{densita5b}
\end{eqnarray}
The Friedel oscillations of the density are then characterized by the wavelength $\lambda^F_e=2 L/ N_\rho$ for even $N_\rho$ (with $N_{\rho}/2$ maxima), and by the superposition of the wavelengths $\lambda^F_{o}=2 L/( N_\rho\pm 1)$ for odd $N_\rho$ (with $(N_{\rho}\pm 1)/2$ maxima). On the other hand the wavelength of Wigner oscillations is always $\lambda^W=L/ N_\rho$ with $N_{\rho}$ maxima.
Fig.~\ref{fig:densita1} shows the typical competition between Friedel and Wigner oscillations induced by the interactions: for stronger interactions (low $g$) Wigner oscillations are enhanced, see panel (a), while at weaker interactions, see panel (b), Friedel oscillations dominate. This behavior is due to the different power laws given in Eqs. (\ref{densita5a}) and (\ref{densita5b})
as a function of the interaction parameter $g$. As $g\to 0$ (strong interactions) the Wigner weight in Eq. (\ref{densita5b}) saturates to one, while the Friedel part in Eq. (\ref{densita5a}) decreases. We want to stress here that we are still in a strong interactions regime, characterized by $v_{\sigma}\ll v_{\rho}$ where the Wigner contribution is always present.

\begin{figure}[htbp]
\begin{center}
\includegraphics[width=12cm,keepaspectratio]{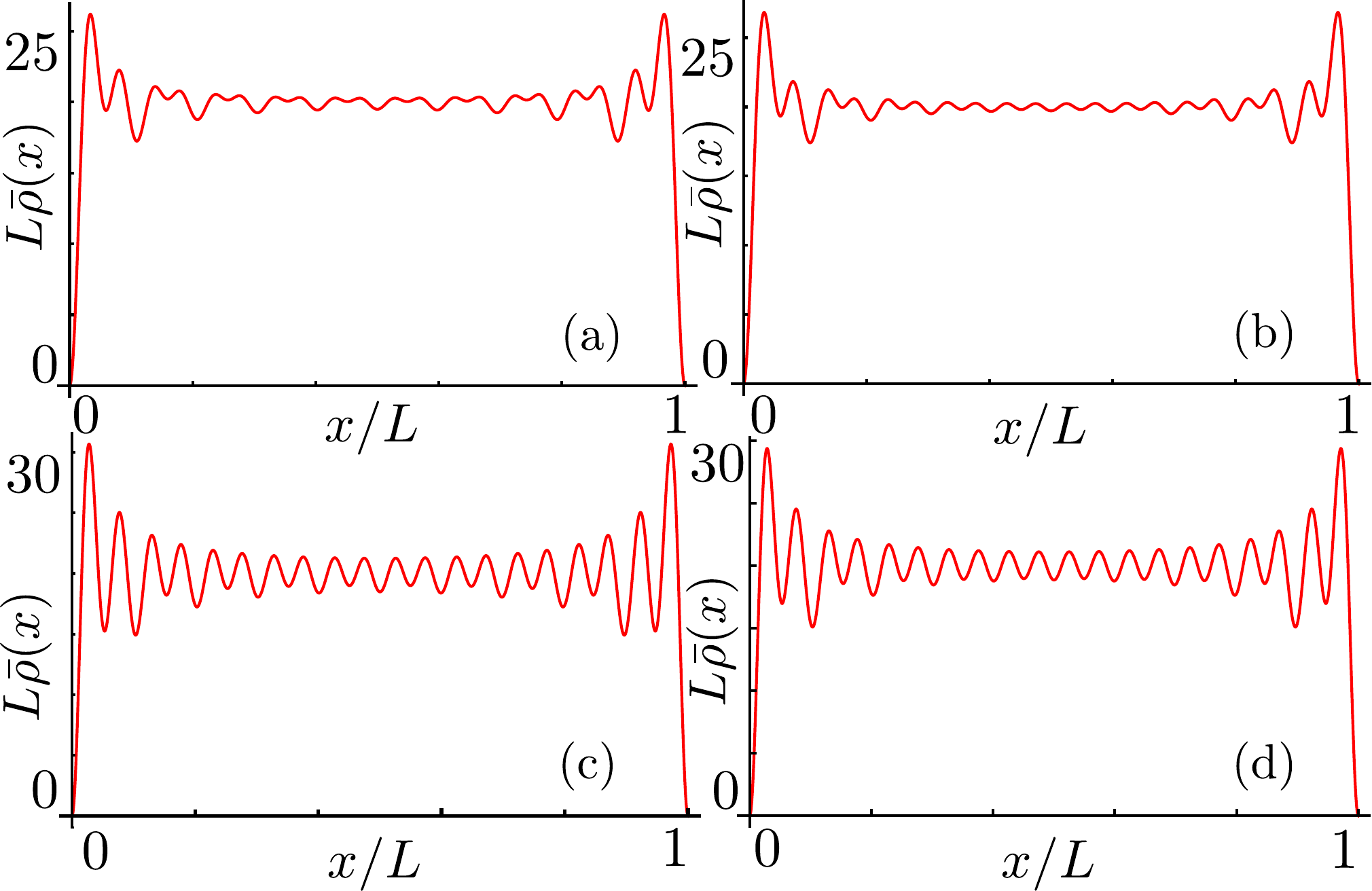}
\caption{(Color online) Adimensionalized electron density as a function of $x/L$ for $N^{(0)}_{\rho}=20$ and
 (a) $g=0.8$, $k_BT=E_\sigma$; (b) $g=0.8$, $k_BT=2E_\sigma$; (c)  $g=0.5$, $k_BT=E_\sigma$; (d) $g=0.5$, $k_BT=2E_\sigma$.
In all panels $v_\rho=40v_\sigma$ and  $\lambda=0.7$.}
\label{fig:densita2}
\end{center}
\end{figure}

It is now natural to consider the stability of the above result against temperature. Indeed, as will be shown below, a tendency towards the emergence of the Wigner contribution over the Friedel one occurs. At finite temperature the averages in Eq.~(\ref{densita4}) are given by the sums
\begin{eqnarray}
\langle\varphi_\rho^2(x)\rangle&=&g\sum_{n>0} \frac{e^{-\frac{\pi\alpha n}{L}}}{n}\sin^2\left(\frac{\pi n x}{L}\right) [2n_B(n\epsilon_\rho,T)+1]\, ,\\
\langle\varphi_s^2(x)\rangle&=&\sum_{\nu=\rho,\sigma}g_{\nu}\sum_{n>0} \frac{e^{-\frac{\pi\alpha n}{L}}}{2n}\sin^2\left(\frac{\pi n x}{L}\right) [2n_B(n\epsilon_\nu,T)+1]\, ,
\end{eqnarray}
with $n_B(\epsilon,T)=[e^{\epsilon/k_BT}-1]^{-1}$.
They will be evaluated numerically. Fig.~\ref{fig:densita2} shows the behavior of the electron density for different temperatures. Increasing temperature a much clearer signature of Wigner oscillations develops: at higher temperatures the Friedel oscillations tend to be washed out leaving the Wigner ones almost unperturbed, which thus acquire a better visibility in comparison to the $T=0$ case discussed in
Fig.~\ref{fig:densita1}. This phenomenon is clearly present both at higher and smaller $g$.
To explain this phenomenon let us remember that at finite temperature ($k_BT\geq E_\sigma$), the Friedel part of the density (Eq.~(\ref{densita2})) contains several contributions given by all the possible excitations of the spin zero modes $N_\sigma$. These terms oscillate with different wavelengths which, for given values $N_{\rho}$ and $N_{\sigma}$, are $\lambda^F_\pm(N_\rho,N_\sigma)=2L/(N_\rho\pm N_\sigma)$.
Their superposition causes a partial cancellation of the Friedel contribution, with a global reduction of its amplitude. Note that this behavior is absent at zero temperature since only the ground spin number values $N_\sigma=0,\pm 1$ are present in that case.

On the other hand, Wigner oscillations are not sensitive to the spin imbalance $N_\sigma$, and hence are not decreased in amplitude. Note that the tendency to the suppression of Friedel oscillations is in accordance with the predictions of the extreme case of spin incoherent Luttinger liquid,~\cite{fiete1} where the spin degree of freedom does not affect the electron density and only Wigner oscillations are present.\\

\subsection{Couplings}
We would now like to consider a possible way to detect the temperature-enhancement effects of Wigner correlations via transport. In particular we would like to analyze how universal is this trend considering transport properties with external probes not necessarily proportional to the electron density. Investigations at low temperatures ($k_{B}T\ll E_{\sigma}$) with an AFM tip were already performed. There, it was demonstrated a direct map between the linear conductance trace and the shape of the electron density~\cite{mantelli,sgm2,AFM}. Here, we would like to consider an STM tip which is not proportional to the density and will give complementary and additional information on this phenomenon~\cite{flensberg,secchi2}.

\begin{figure}[htbp]
\begin{center}
\includegraphics[width=8.8cm,keepaspectratio]{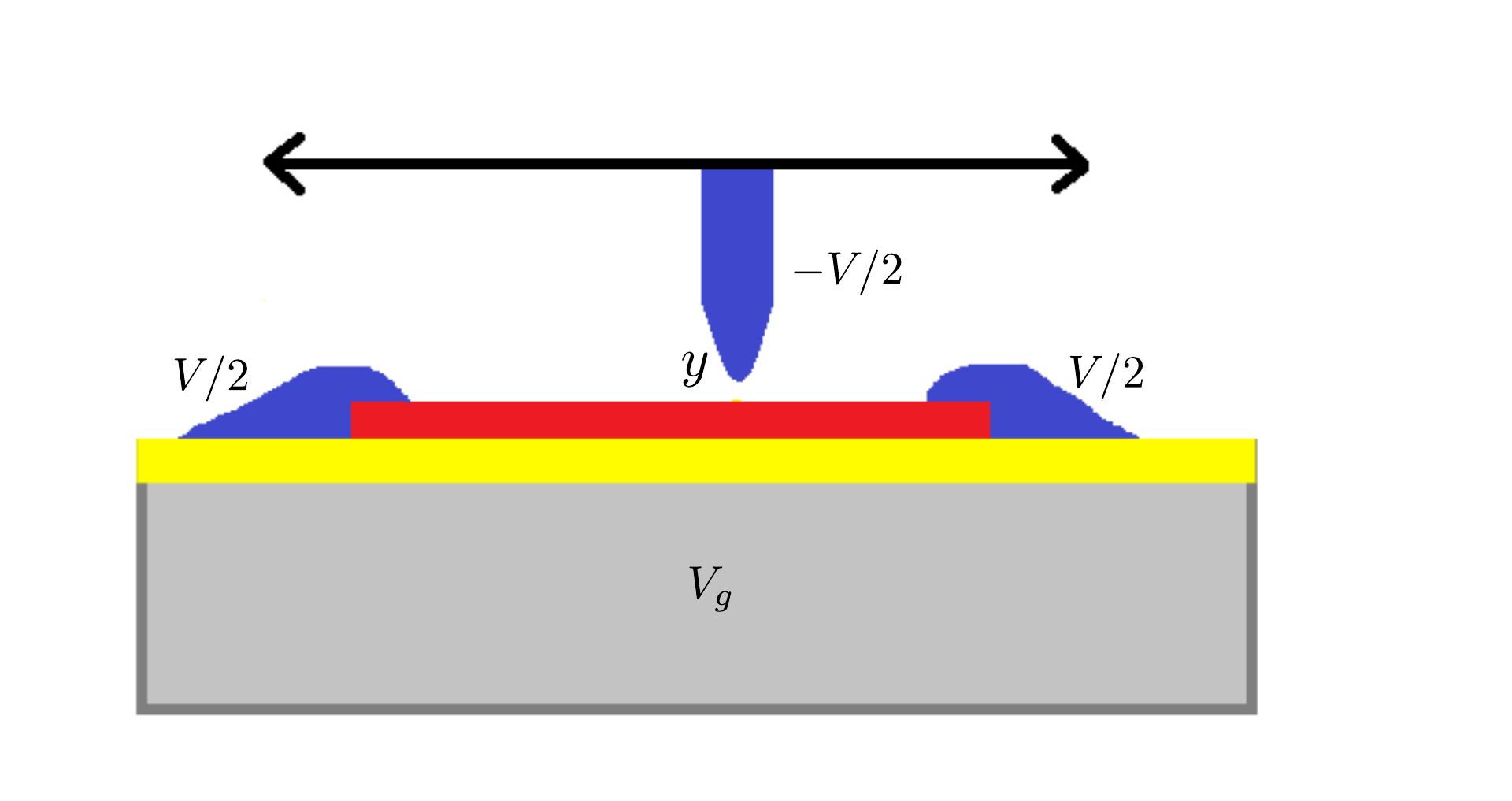}
\caption{(Color online) Schematic setup of the 1D quantum dot (red), connected to the STM tip at position $y$ and voltage $-V/2$ and to the lateral contacts at voltage $V/2$. The gate is at voltage $V_g$.}
\label{fig:figsetup}
\end{center}
\end{figure}

\noindent As sketched in Fig.~\ref{fig:figsetup} the STM tip is coupled to the quantum dot and is free to move along the $x$-axis. In addition, standard tunneling couplings with left and right leads at positions $x_1=0$ and $x_2=L$ are considered. The tip is at potential $-V/2$, while the lateral contacts are both at potential $V/2$. The gate, capacitively coupled to the dot, is at potential $V_g$. Voltage drops are assumed to occur symmetrically on the dot.

\noindent The non-interacting leads ($H^{leads}$) are connected to the quantum dot via standard tunnel barriers
\begin{equation}
{H}_{t}^{leads}=t_{0}\sum_{j=1,2}\sum_{s}{\psi}_{s,+}(x_{j})\chi_{s,j}^\dag(x_{j})\!+\!\mathrm{h.c.}\,,\end{equation}
with $t_{0}$ the tunneling amplitude and ${\chi}_{s,j}(x)$ the electron lead operators with spin $s$ and $j=1$ or $j=2$ for right or left contacts. \\
The STM tip is modeled as a semi-infinite Fermi contact, with Hamiltonian $H^{tip}$,
placed above the dot at a position $y$. The tunnel coupling $H_t^{tip}$ is expressed in terms of the tip Fermi field operator $\psi_{s,F}(z)$ ($z$ is the coordinate
along the tip with $z = 0$ at the vertex),~\cite{bercioux} with tunneling barrier transparency $\tau$
\begin{eqnarray}
H_{t}^{tip}&\!=\!&\tau\sum_s \psi_{s,F}(0^+)\left[ h_{t,s}^{(0)}+L\ c\ h_{t,s}^{(1)}\right]\!+\!\mathrm{h.c.},\label{eq:tuntip}\\
h_{t,s}^{(0)}&=&e^{-ik_Fx}\psi^{\dagger}_{s,+}(x)+e^{ik_Fx}\psi^{\dagger}_{s,-}(x)\,, \nonumber\\
h_{t,s}^{(1)}&=&\sum_{r=\pm}e^{-3rik_Fy}\psi^\dag_{s,r}(y)\psi^\dag_{-s,r}(y)\psi_{-s,-r}(y)\,.\nonumber
\end{eqnarray}
The above Hamiltonian consists of two different terms: the standard one ($h^{(0)}_{t,s}$), which represents injection into the dot of an electron of spin $s$ and both chiralities and an additional term ($h_{t,s}^{(1)}$) which describes the tunneling of an electron of spin $s$ and chirality $r$ together with an electron-hole pair backscattering with opposite spin. Note that this process is the first term of a series involving higher order many body tunneling events~\cite{tunneling,safi} and is expected to be nonzero only in the presence of electron interactions. The dimensionless factor $c$ measures the weight of this term, with $c=0$ in the absence of interactions. In the following, we will consider it as a free parameter.

\section{Transport}
\label{sec:transport}

In the rest of the paper we will focus on the sequential transport regime, treating the tunnel barriers (leads and tip) at lowest order.
The bosonic charge and spin density wave excitations are assumed to be relaxed at thermal equilibrium due to possible couplings with external perturbations~\cite{ioale1}.
The only degrees of freedom out of equilibrium are then the zero charge and spin modes. They must be explicitly retained in the stationary rate equation for the dot occupation probability $P_{\mathcal{S}}$ of a generic dot state $|{\mathcal S}\rangle\equiv|N_{\rho},N_{\sigma}\rangle$:
$\sum_{\mathcal{S}'\neq\mathcal{S}}\left[P_{\mathcal{S}'}\Gamma_{\mathcal{S}'\to\mathcal{S}}-P_{\mathcal{S}}\Gamma_{\mathcal{S}\to\mathcal{S}'}\right]=0$.
Here, $\Gamma_{\mathcal{S}\to\mathcal{S}'}$ are the tunneling rates between the states  $|{\mathcal S}\rangle $ and $|{\mathcal S}'\rangle$. They consist of two contributions coming from the two tunneling Hamiltonians
\begin{equation}
\Gamma_{\mathcal{S}\to\mathcal{S}'}=\Gamma_{\mathcal{S}\to\mathcal{S}'}^{leads}+\Gamma_{\mathcal{S}\to\mathcal{S}'}^{tip}(y)\, .\label{eq:rategen}
\end{equation}

\noindent The explicit expression of these terms can be derived via the time evolution of the density operator. In the interaction picture with respect to
$H^{(0)}=H_{d}+H^{leads}+H^{tip}$ one has ~\cite{blum,nonlin1}
\begin{equation}
\rho(t)=Te^{-i\int_0^{t} dt' {\cal V}(t')}\rho(0)\widetilde Te^{i\int_0^{t} dt' {\cal V}(t')}\, ,
\label{densitatempo}
\end{equation}
with ${\mathcal V}(\tau)=e^{iH^{(0)}\tau}(H_{t}^{tip}+{H}_{t}^{leads})e^{-iH^{(0)}\tau}$
and $T$ ($\widetilde T$) the time (anti-time) ordering. As discussed above, the initial density operator is chosen as
\begin{equation}
\rho(0)=\frac{e^{-\beta \left(H^{leads}+H^{tip}\right)}}{Z_{leads}Z_{tip}}\frac{e^{-\beta H_b}}{Z_{b}}|\mathcal{S}\rangle\langle\mathcal{S}|\, ,
\end{equation}
with leads, tip, spin and charge density wave excitations in their own thermal equilibrium
with the partition functions
$Z_r=Tr\left\{e^{-\beta H_r}\right\}$ ($r\in\{{b,leads,tip}\}$).

\noindent Following a standard expansion procedure~\cite{AFM} at lowest order of ${\mathcal V}(t)$ in Eq. (\ref{densitatempo}), one gets
 ($\eta= {leads, tip}$)
\begin{equation}
\Gamma^{\eta}_{\mathcal{S}\to\mathcal{S}'}=\lim_{t\to\infty}\int_0^{t}d{t'}[F^{\eta}(t',t)+F^{\eta}(t,t')]\,,
\label{rate1}
\end{equation}
with
\begin{equation}
F^{\eta}(t_1,t_2)=Tr\langle \mathcal{S}'|H_t^{\eta}(t_1)\rho(0)H_t^ {\eta}(t_2)|\mathcal{S}'\rangle.\label{eq:F0}\end{equation}
The explicit calculation of Eq. (\ref{rate1}) can be carried out using the bosonization formalism~\cite{AFM}. In Appendix~\ref{sec:appa} the evaluation of $\Gamma^{tip}_{\mathcal{S}\to\mathcal{S}'}$ is presented.
In the following, we will solve the master equation in the linear regime for $k_{B}T\ll E_\rho$. This ensures that the sequential transport involves two charge states only, say $N_\rho=N_1$ and $N_\rho=N_2=N_1+1$.
The possible states $|\mathcal{S}\rangle$ that enter the master equation are then characterized by all possible spin imbalance numbers
\begin{equation}
|\mathcal{S}\rangle=|N_{\rho},N_{\sigma}\rangle= |N_l,N_l-2j\rangle\equiv|\mathcal{S}(l,j)\rangle \end{equation}
with $l=1,2$, and $0\leq j\leq N_l$.
In addition, in order to emphasize the effects of the tunneling from the tip, we will consider the barrier tip much higher than the one of the leads ($\tau\ll t_0$). In this case the solution
for the linear conductance ${G}$ near resonance simplifies to~\cite{master}
\begin{equation}
{G}=\frac{e^2}{k_BT}\sum_{j,j'} P_0^{j}\Gamma^{tip}_{\mathcal{S}(1,j)\rightarrow\mathcal{S}(2,j')}(y)\,, \label{eq:furu}\end{equation}
with  the equilibrium probabilities
\begin{eqnarray}
P_0^j&=& e^{-{\beta\mathcal{E}_{1,j}}}/\sum_{j,l}e^{-{\beta\mathcal{E}_{l,j}}},\\
\mathcal{E}_{l,j}&=&\langle \mathcal{S}(l,j)|H_{\rho}+H_{\sigma}| \mathcal{S}(l,j)\rangle.
\end{eqnarray}
Note that since we examine the sequential tunneling regime, the selection rule $|2j+1-2j'|=1$ holds. In order to obtain the result in Eq.~(\ref{eq:furu}) detailed balance has been used, linking the rates $\Gamma^{tip}_{\mathcal{S}(2,j')\rightarrow\mathcal{S}(1,j)}(y)$ with $\Gamma^{tip}_{\mathcal{S}(1,j)\rightarrow\mathcal{S}(2,j')}(y)$.
\subsection{Results}
\begin{figure}[htbp]
\begin{center}
\includegraphics[width=12cm,keepaspectratio]{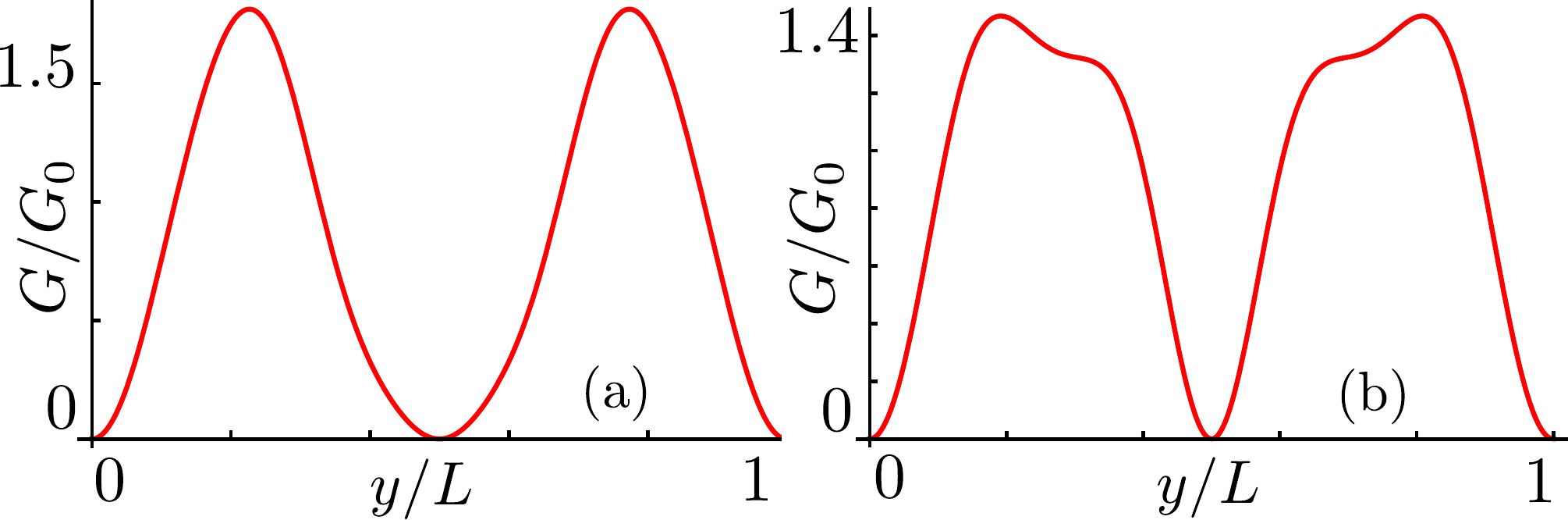}
\caption{(Color online) Linear conductance, normalized to $G_0$, at low temperatures (Eq.~(\ref{eq:conduttanza1})) as a function of $y/L$ with $c=0.3$ and $g=0.5$ for (a) $N_1=2$, $N_2=3$, and (b) $N_1=3$, $N_2=4$.}
\label{fig:linront}
\end{center}
\end{figure}

We begin the discussion considering low temperatures $k_BT\ll E_\sigma$. In this regime
only transitions between ground states  characterized by $N_1$ and $N_2=N_1+1$ electrons are relevant.
Plugging the low temperature limit for the tip tunneling rate - Eq.~(\ref{eq:rate}) - into Eq.~(\ref{eq:furu}), one finds an expression for the on resonance linear conductance ${G}$. One has
\begin{equation} 
G=G_{0}d_0(y)^{\frac{1-g^2}{8g}}\left\{r_1(y)+c[r_2(y)+r_3(y)]+c^2r_4(y)\right\}\label{eq:conduttanza1}\, ,
\end{equation}
with
\begin{eqnarray} 
G_0&=&\frac{e^2|\tau|^2\nu_{tip}}{4\sqrt{2}\alpha k_BT\cosh^2\left(\frac{\ln2}{4}\right)}(1-e^{-\alpha\pi/L})^{\frac{1+g}{2}},\nonumber\\
d_{0}(y)&=&\left[\frac{\sinh\left(\frac{\pi\alpha}{L}\right)}{\sqrt{\sinh^2\left(\frac{\pi\alpha}{L}\right)+\sin^2\left(\frac{\pi y}{L}\right)}}\right]^{-2},\nonumber\\
r_1(y)&=&1-\cos\left[\frac{2\pi y}{L}\left(\frac{N_2}{2}+\frac{1}{4}\pm\frac{1}{4}\right)\right],\nonumber\\
r_2(y)&=&-\frac{d_{0}(y)^{-\frac{g}{2}}}{\pi \alpha}\cos\left[\frac{2\pi y}{L}N_{2}-2h(y)\right],\nonumber\\
r_3(y)&=&\frac{d_{0}(y)^{-\frac{g}{2}}}{\pi \alpha}\cos\left[\frac{2\pi y}{L}\left(\frac{N_2}{2}-\frac{1}{4}\mp\frac{1}{4}\right)-2h(y)\right],\nonumber\\
r_4(y)&=&\frac{d_{0}(y)^{-g}}{4\pi^2 \alpha^2}\left\{1-\cos\left[\frac{2\pi y}{L}\left(\frac{3N_2}{2}-\frac{1}{4}\mp\frac{1}{4}\right)-4h(y)\right]\right\}\, .\nonumber
\end{eqnarray}
In the above expressions the sign $\pm$ stands for even/odd $N_1$, $\nu_{tip}$ is the density of states per unit volume in the tip, and $h(x)$ is defined in Eq.~(\ref{eq:effe}).
In this expression contributions with different wavelengths are present. For even/odd $N_1$, the oscillations associated to the  contributions $r_i(y)$  ($i=1,..,4$) are characterized by the wavelengths $\lambda^{(r_1)}=L/\left({N_2}/{2}+{1}/{4}\pm/{1}/{4}\right)$,
$\lambda^{(r_2)}=L/\left({N_2}\right)$,
$\lambda^{(r_3)}=L/\left({N_2}/{2}-{1}/{4}\mp{1}/{4}\right)$, and
$\lambda^{(r_4)}=L/\left({3N_2}/{2}-{1}/{4}\mp{1}/{4}\right)$.\\
Note that the Friedel oscillations are given by the  $r_1(y)$ term, while $r_2(y)$ represents the Wigner oscillations. In addition, other  contributions are present: indeed the relation in Eq.~(\ref{eq:conduttanza1}) clearly shows that the linear conductance is not simply  proportional to the local electron density, and represents an alternative playground for the characterization of Wigner correlations at finite temperature.\\
Figure~\ref{fig:linront} shows the low temperature conductance as a function of the tip position at low particle numbers. Even in the presence of relatively strong interactions $g=0.5$, the resonances  between ($N_1=2$; $N_2=3$)  panel (a) and ($N_1=3$; $N_2=4$)  panel (b) do not show clear evidences of Wigner oscillations. The latter should indeed have three and four maxima respectively. Note that this behavior with weight  $c=0.3$ between Friedel and Wigner  contributions
is in good agreement with  the numerical results obtained by Secchi and Rontani~\cite{secchi2} at low particle numbers.

\begin{figure}[htbp]
\begin{center}
\includegraphics[width=12cm,keepaspectratio]{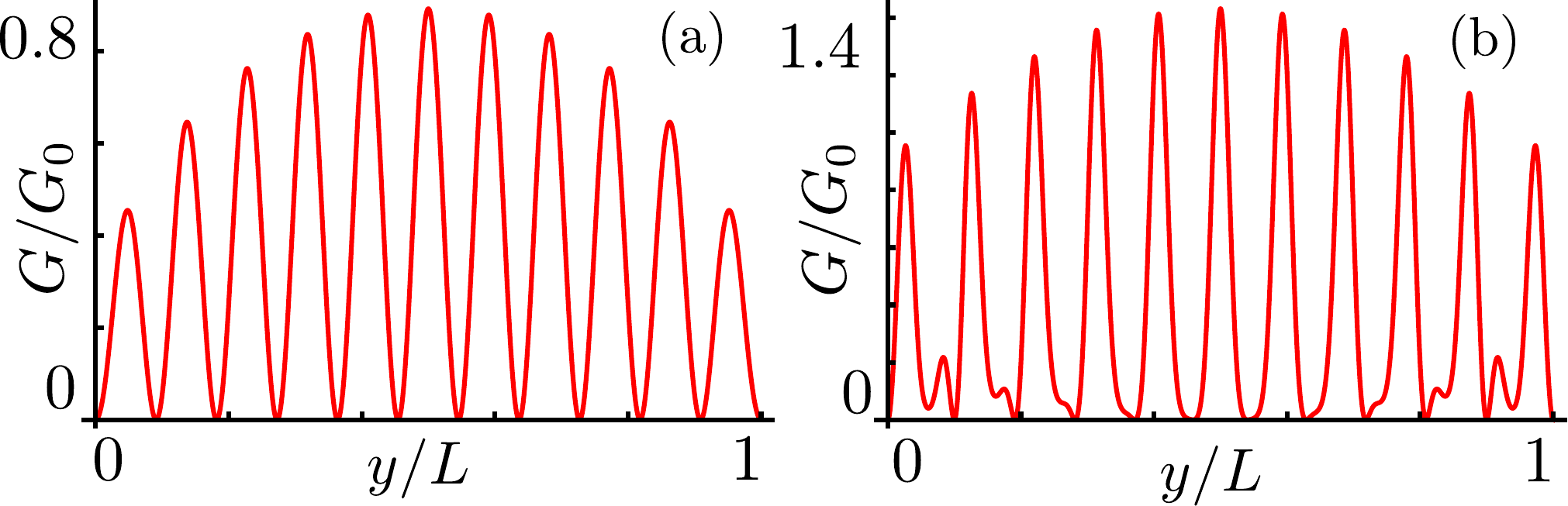}
\caption{(Color online) Linear conductance as in Fig.~\ref{fig:linront} but for $N_{1}=20$ and $N_{2}=21$ electrons and (a) $c=0$; (b) $c=0.3$.}
\label{fig:linear}
\end{center}
\end{figure}
\noindent The behavior at higher electron numbers can be also checked within our approach. As shown in Fig.~\ref{fig:linear}(b), in the presence of the same interaction strength as in Fig.~\ref{fig:linront},
Wigner oscillations are also absent (they should have 21 maxima), and the conductance shows a characteristic Friedel oscillation. For comparison in panel (a) we show the pure Friedel contribution
obtained considering the $c=0$ value in the conductance expression.

\noindent The Luttinger liquid theory allows to explain this behavior: the power law governing the dependence of the low temperature conductance with respect to  interactions does not favor the Wigner oscillations, even in the $g\rightarrow 0$ limit, where all oscillations scale as $1/(8g)$ (see Eq.~(\ref
{eq:conduttanza1})). Note that the result is not in contradiction to what previously found for the electron density, where Wigner oscillations are enhanced increasing interactions since the STM tip is {\em not} directly coupled to the density.\\

Let us now investigate the behavior at larger temperatures. In this regime, the relation in Eq.~(\ref{eq:furu}) has to be employed. It consists of
several contributing rates corresponding to different spin excited values. They will start to play for temperatures above the spin addition energy $E_{\sigma}$. In particular, we expect that they will affect the Friedel term, which indeed depends on the total number of excited electrons \textit{per spin}, producing higher harmonics with different amplitudes with respect to the one identified at $T=0$.
We will see that the superposition of these oscillations will produce partial cancellations, masking the visibility of the subperiods and also decreasing the overall principal Friedel oscillations.\\

\begin{figure}[htbp]
\begin{center}
\includegraphics[width=12cm,keepaspectratio]{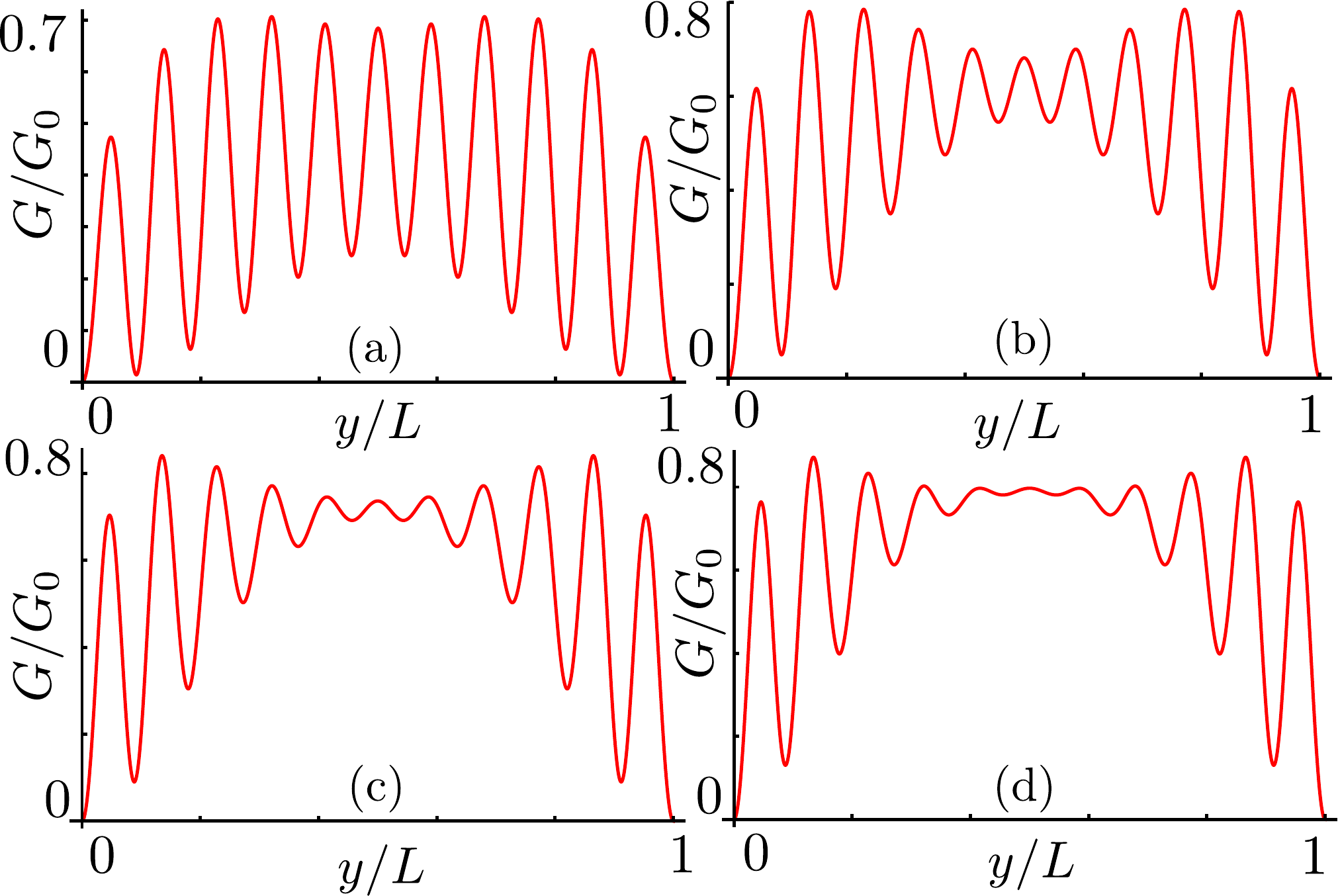}
\caption{(Color online) Linear conductance, normalized to $G_0$, for $N_{1}=20$ and $N_{2}=21$ electrons as a function of $y/L$ and $g=0.5$ for the Friedel contribution only ($c=0$). (a) $k_BT=E_\sigma$; (b) $k_BT=2E_\sigma$; (c) $k_BT=3E_\sigma$; (d) $k_BT=4E_\sigma$.}
\label{fig:solofriedel}
\end{center}
\end{figure}

In order to show this fact we start by considering the Friedel contribution {\em alone} with $c=0$, where only long wave and Friedel oscillations are retained in the conductance. The corresponding conductance is shown in Fig.~\ref{fig:solofriedel}. Raising the temperature the amplitude of the main oscillations decreases, without the appearance of new peaks.
Apparently the weights associated to the higher harmonics are not strong enough to allow them to appear as new maxima. However, due to interference they tend to suppress the amplitude of the primary oscillating mode.\\
This behavior is analogous to the one characterizing the electron density.
In addition, the nodes of the conductance are now located at the extremal points of the quantum dot, inner nodes present at zero temperature disappear due to the long wave contribution to the tunneling rates activated at finite temperature. \\
The temperature behavior will be completely different concerning Wigner oscillations. Indeed, since the Wigner term depends on the total number of electrons only, it will be mainly unaffected by the different spin excited numbers. In other words, no additional higher harmonics will be present for the Wigner part. The number of maxima will always be given by $N_{2}$ (the number of electrons in the final state). They will not be suppressed by cancellations even increasing temperatures.
\begin{figure}[htbp]
\begin{center}
\includegraphics[width=12cm,keepaspectratio]{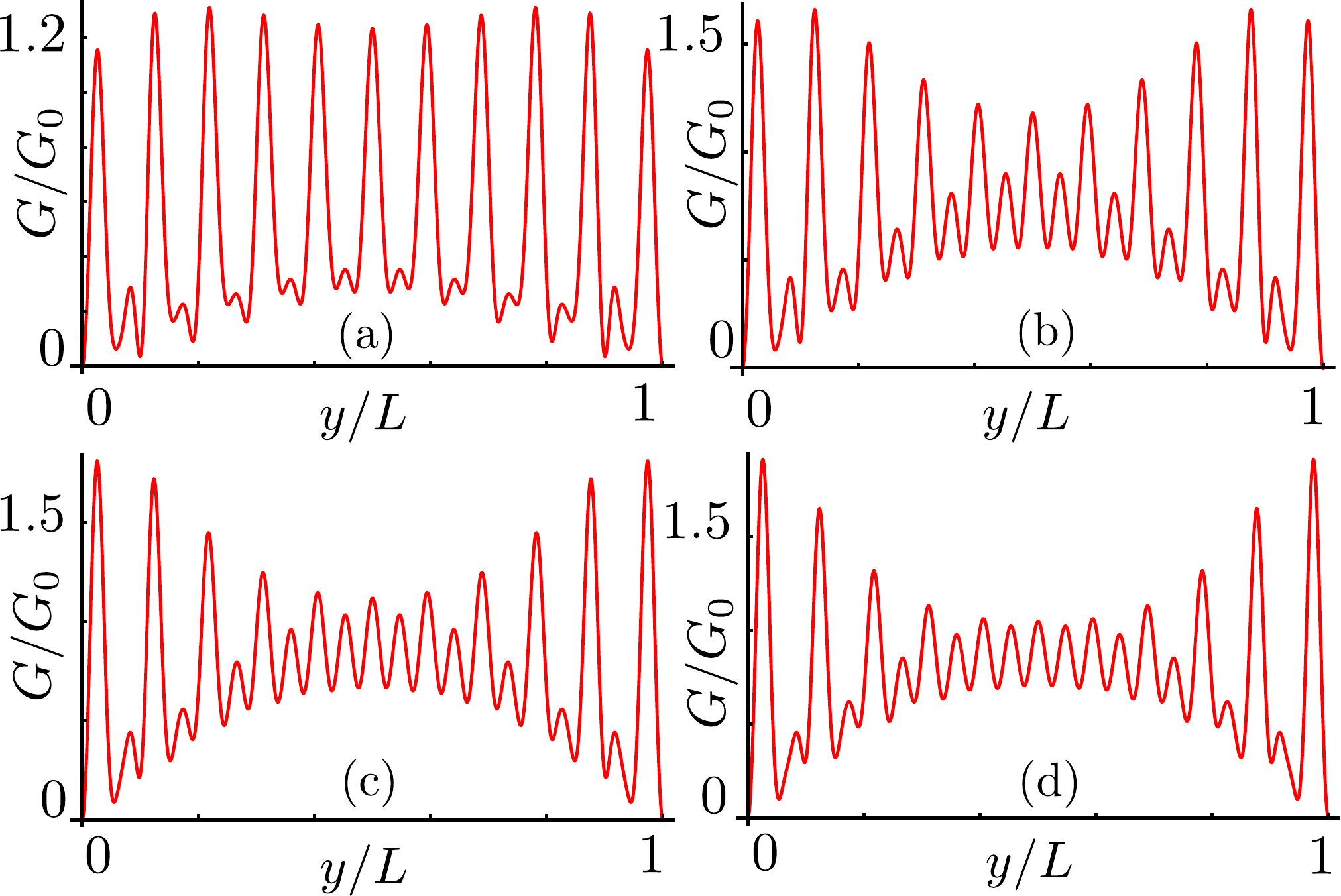}
\caption{(Color online) Linear conductance, normalized to $G_0$, for $N_{1}=20$ and $N_{2}=21$ electrons as a function of $y/L$ for: (a) $T=k_BE_\sigma$; (b) $k_BT=2E_\sigma$; (c) $k_BT=3E_\sigma$; (d) $k_BT=4E_\sigma$. Other parameters $c=0.3$, $g=0.5$.}
\label{fig:temp}
\end{center}
\end{figure}

\noindent Figure~\ref{fig:temp} shows the overall behavior of the conductance. Raising  temperature the Friedel oscillations decrease and Wigner peaks start to emerge. Already for $k_{B}T\sim4E_\sigma$ Wigner type oscillations become clearly visible at the expenses of the Friedel term. Note that this trend is more pronounced in the central part of the dot.

We would like now to conclude this part with some comment. All the results are obtained within the coherent Luttinger liquid picture.  This means that the condition $k_BT<D_\sigma=N_\rho E_\sigma$ must always be fulfilled. Indeed, this constraint can be satisfied in our calculations, adjusting the number $N_\rho$ of electrons in the dot. For low numbers and high temperatures other techniques should be employed.\\
A second comment is about the energy scales of excitations. The cancellation effects in the Friedel channel, induced by temperature, are driven by the spin addition energy $E_{\sigma}$.
Wigner oscillations, on the other hands, are insensitive to this scale and stable, since they are connected to the charge addition energy $E_{\rho}$ only.
Consequently, at strong enough interactions where $E_\sigma\ll E_\rho$ we expect a finite temperature range $E_\sigma<k_BT\ll E_\rho$ where Wigner correlations are dominant and more easily detectable. Only at even higher temperatures $k_BT>E_{\rho}$, where the quantum dot contains different particle numbers (not considered here)  Wigner oscillation will be also depressed.\\
Finally it is worth noticing that the role of the spin density waves, that are automatically included in the tunneling rates at finite temperature, see Eq.~(\ref{eq:rate}), is to strengthen the suppression of the Friedel oscillations.\\
Finally, a brief comment on the non-linear transport regime for $eV>k_{B}T$. Indeed, even at low temperature $k_{B}T\ll E_{\sigma}$ but large enough voltage $eV>E_{\sigma}$ we expect a similar suppression of the Friedel signal in comparison to the Wigner one, due to the stochastic population of transport channels involving different spin zero modes. This issue is being analyzed at the moment.

\section{Conclusions}
\label{sec:concl}
We have studied the temperature-induced emergence of Wigner correlations over finite-size effects in a strongly interacting 1D quantum dot, within the coherent Luttinger liquid picture. We demonstrated that, for  temperatures comparable with  the zero mode spin excitations, Friedel oscillations are suppressed by the cancellations due to superposition of oscillations with several different subperiods: Wigner oscillations are consequently enhanced and stable.\\
\noindent This effect is shown to be general and occurs both in the electron density and in the linear conductance in the presence of an STM tip. The evidence brought by this latter probe is of particular significance since it is not directly affected by the electron density, thus bringing complementary, independent information with respect to the latter.\\
\noindent For an experimental observation of the above predictions, a clean electron systems is very important since disorder may pin the Wigner molecule distorting its shape, as well as directly influence the transport properties of the channel. In this sense. possible candidates could be semiconducting systems~\cite{STMWIRE} which however are rather difficult to be investigated with STM probes due to a low density of states. Carbon nanotubes are other notable systems which should exhibit the effects discussed above, being characterized by high mobilities and being well suited to the investigation by means of an STM~\cite{STMCNT} The extension of the above theory to the case of a carbon nanotube will be the subject of future work.

\noindent\textit{Acknowledgments.} Financial support by the EU-FP7 via ITN-2008-234970 NANOCTM is gratefully acknowledged.
\appendix
\section{Tunneling rates}
\label{sec:appa}
The tunneling rates needed for the evaluation of the sequential linear conductance in Eq.~(\ref{eq:furu}) are $\Gamma_{\mathcal{S}\to\mathcal{S}'}^{tip}$, with states $|\mathcal{S}\rangle=|N_\rho,N_\sigma\rangle$ and $|\mathcal{S}'\rangle=|N_\rho+1,N_\sigma'\rangle$, and $|N_\sigma-N_\sigma'|=1$. Substituting  Eq.~(\ref{eq:tuntip}) in Eq.~(\ref{eq:F0}) for the tip part leads to
\begin{eqnarray}
F^{tip}(t_{1},t_{2},y)&=&{|\tau|^2}e^{-W_{tip}(t_2-t_1)}\mathcal{F}(t_1,t_2,y),\nonumber\\
\mathcal{F}(t_1,t_2,y)&=&\sum_{i=1}^6 Tr \langle\mathcal{S}'| \mathcal{F}^{tip,i}(t_{1},t_{2},y)|\mathcal{S}'\rangle,
\label{A1}
\end{eqnarray}
where the explicit dependence on the tip position $y$ has been reintroduced. The kernel $W_{tip}(\tau)$, due to the thermal average on the tip degrees of freedom, is~\cite{ioale1,AFM,nonlin2}
\begin{equation}ß
e^{-W_{tip}(\tau)}=\nu_{tip}\int_{-\infty}^{\infty}\ dE\ e^{iE\tau}f(E)\, ,
\end{equation}
with $\nu_{tip}$ density of states of the tip and $f(E)=1/(1+e^{\beta E})$
the Fermi function. The trace in Eq.~(\ref{A1}) is now only over spin and charge density  wave modes. The different terms in $F^{tip}(t_{1},t_{2},y)$ are
\begin{eqnarray}
\mathcal{F}^{tip,1}(t_{1},t_{2},y)&=&\sum_{r,s}\psi^{\dag}_{s,r}(y,t_1)\rho(0)\psi_{s,r}(y,t_2),\nonumber\\
\mathcal{F}^{tip,2}(t_{1},t_{2},y)&=&\sum_{r,s}e^{2irk_Fy}\psi^{\dag}_{s,-r}(y,t_1)\rho(0)\psi_{s,r}(y,t_2),\nonumber\\
\mathcal{F}^{tip,3}(t_{1},t_{2},y)&=&c\sum_{r,s}e^{irk_Fy}\left[h^{(1)}_{t,s,-r}(t_1)\rho(0)\psi_{s,r}(y,t_2)+\psi^\dag_{s,-r}(y,t_1)\rho(0)h^{(1)\dag}_{t,s,r}(t_2)\right],\nonumber\\
\mathcal{F}^{tip,4}(t_{1},t_{2},y)&=&c\sum_{r,s}e^{irk_Fy}\left[h^{(1)}_{t,s,r}(t_1)\rho(0)\psi_{s,r}(y,t_2)+\psi^\dag_{s,-r}(y,t_1)\rho(0)h^{(1)\dag}_{t,s,-r}(t_2)\right],\nonumber\\
\mathcal{F}^{tip,5}(t_{1},t_{2},y)&=&c^2\sum_{r,s}h^{(1)}_{t,s,r}(t_1)\rho(0)h^{(1)\dag}_{t,s,r}(t_2),\nonumber\\
\mathcal{F}^{tip,6}(t_{1},t_{2},y)&=&c^2\sum_{r,s}h^{(1)}_{t,s,r}(t_1)\rho(0)h^{(1)\dag}_{t,s,-r}(t_2).\nonumber
\end{eqnarray}
In order to evaluate the trace over spin and charge modes, the bosonized form of the fermion operators $\psi_{s,+}(x)$, given in Eq.~(\ref{eq:opright}) is used, together with the relation $\psi_{s,r}(x)=-\psi_{s,-r}(-x)$. Here, we quote the result for the first of these terms, the contributions arising from the others are similar and will not be reported
%\begin{widetext}
%\begin{align}
\begin{eqnarray}
Tr\langle\mathcal{S}'| \mathcal{F}^{tip,1}(t_{1},t_{2},y)|\mathcal{S}'\rangle&=&\frac{e^{-i\Delta E^{tip}\Delta t}}{2\pi\alpha}d(y)
e^{-\left[\alpha_{g}^{+}W(\epsilon_\rho,\Delta t)+\sum_{p=\pm}\frac{\alpha_{g}^{-}}{2}W\left(\epsilon_\rho,\Delta t+\frac{py}{gv_\rho}\right)+\frac{1}{2}W(\epsilon_\sigma,\Delta t)\right]}\nonumber\\
&+&\mathrm{h.c.}\nonumber
\end{eqnarray}
%\end{align}
%\end{widetext}
where $\alpha_{g}^{\pm}=(1/4g)\pm(g/4)$. Here, the kernels $W{(\epsilon,\tau)}$ arise from the thermal average over the spin and charge waves of the dot and read as~\cite{ioale1}
\begin{equation}
W(\varepsilon,\tau)=\sum_{n>0}\frac{e^{-\pi\alpha n/L}}{n}\left\{\coth{\left(\frac{\beta n\varepsilon}{2}\right)}\left[1-\cos(n\varepsilon\tau)\right]+i\sin(n\varepsilon\tau)\right\}\, .\label{eq:W}
\end{equation}
In addition, we have defined
\begin{eqnarray}
 \Delta t&=&t_2-t_1\nonumber\\
\Delta E^{tip}&=&-\frac{eV}{2}+\Delta E_0,\nonumber\\
\Delta E_0&=&E_0(N_\rho+1,N'_\sigma)-E_0(N_\rho,N_\sigma),\nonumber\\
E_0(N_\rho,N_\sigma)&=&\frac{E_\rho}{2}(N_\rho-N_\rho^{(0)}-N_g)^2+\frac{E_\sigma}{2}N_\sigma^2;\nonumber\end{eqnarray}
with
\begin{equation} d(y)=\left|e^{W\left(\epsilon_\rho,{gy}/{v_\rho}\right)}\right|^2.\end{equation}
Note that in the low temperature limit $d(y)\to d_{0}(y)$, as given in Eq.~(\ref{eq:conduttanza1}), with
\begin{equation} d_{0}(y)=\left[\frac{\sinh\left(\frac{\pi\alpha}{L}\right)}{\sqrt{\sinh^2\left(\frac{\pi\alpha}{L}\right)+\sin^2\left(\frac{\pi x}{L}\right)}}\right]^{-2}.\end{equation}
The integral in Eq.~(\ref{rate1}) can now be performed, since $W(\epsilon,\tau)$ is periodic, and can be expressed in the Fourier series ($\kappa >0$)
\begin{equation}
e^{\pm\kappa W(\epsilon,\tau)}=\sum_{l=-\infty}^{\infty}w_{\pm,l}^{\kappa}e^{-il\epsilon\tau}\, .
\end{equation}
For $k_{B}T\ll\varepsilon_{\sigma}$ one can approximate $w_{_\pm,l}^{\kappa}$ with their expressions calculated for $T=0$~\cite{ioale1}
\begin{eqnarray}
\!\!w_{+,l}^{\kappa}\!&=&\!\left(-e^{-\frac{l\pi\alpha}{L}}\right)^{l}\!\left(1-e^{-\frac{\alpha\pi}{L}}\right)^{-\kappa}\!\frac{\Gamma(1+\kappa)\theta(l)}{l!\Gamma(1+\kappa-l)}\\
\!\!w_{-,l}^{\kappa}\!&=&\!\left(e^{-\frac{l\pi\alpha}{L}}\right)^{l}\!\left(1-e^{-\frac{\alpha\pi}{L}}\right)^{\kappa}\!\frac{\Gamma(l+\kappa)}{l!\Gamma(\kappa)}\theta(l),
\end{eqnarray}
where $\Gamma(x)$ is the Euler gamma function.
At higher temperature the weights have been evaluated numerically.\\
Using the Fourier expansions the explicit expression for the tunneling rate $\Gamma_{\mathcal{S}\to\mathcal{S}'}^{tip}$ reads as
%\begin{equation} \Gamma_{\mathcal{S}\to\mathcal{S}'}^{tip}=\label{eq:rate} \end{equation}
\begin{eqnarray}
\frac{\Gamma_{\mathcal{S}\to\mathcal{S}'}^{tip}}{\Gamma_0^{tip}}&=&\sum_{\underline{m}}f(\Delta E^{tip}+\eta_{\underline{m}})
\left\{d(y)^{\frac{1}{8g}-\frac{g}{8}}\left[b_{\underline{m}}^{(1)}\cos\left(\frac{2\pi y(m_2-m_3)}{L}\right)\right.\right.\nonumber\\
&-&\left.\left.b_{\underline{m}}^{(2)}
\cos\left(\frac{2\pi y(N_a+1+(m_2-m_3+m_4))}{L}\right)\right]\right.\nonumber\\
&+&\left.\right.\left.\frac{c d(y)^{\frac{1}{8g}-\frac{5}{8}}}{\pi\alpha}\left[b_{\underline{m}}^{(3)}\cos\left(\frac{2\pi y(N_\rho+1+(m_1-m_2))}{L}-2h(y)\right)\right.\right.\nonumber\\
&-&\left.\left.b_{\underline{m}}^{(4)}
\cos\left(\frac{2\pi y(N_b+(m_1-m_2-m_4))}{L}-2h(y)\right)\right]\right.\nonumber\\
&+&\left.\right.\left.\frac{c^2d(y)^{\frac{1}{8g}-\frac{9g}{8}}}{\pi^2\alpha^2}\left[b_{\underline{m}}^{(5)}\cos\left(\frac{2\pi y(m_2-m_3)}{L}\right)\right.\right.\nonumber\\
&-&\left.\left.b_{\underline{m}}^{(6)}
\cos\left(\frac{2\pi y (N_a+1+2N_b+(m_2-m_3-m_4))}{L}-4h(y)\right)\right]
\right\}.\nonumber
\label{eq:rate}
\end{eqnarray}
Here, $\Gamma_0^{tip}=\nu_{tip}|\tau|^2/\alpha$ and
\begin{equation}
N_{a/b}=N_\rho\pm\bar{s}N_\sigma;\quad\bar{s}=N'_\sigma-N_\sigma,
\end{equation}
with $\underline{m}=(m_{1},m_{2},m_{3},m_{4})$ a four component vector of integers $m_{i}$ and $\eta_{\underline{m}}$ given by
\begin{equation}\eta_{\underline{m}}=\epsilon_\rho(m_1+m_2+m_3)+\epsilon_\sigma m_4.\end{equation}
The weights $b^{(\mu)}_{\underline{m}}$, with $\mu=1,..,6$ are
\begin{eqnarray}
b_{\underline{m}}^{(1)}&=&w_{-,m_1}^{\alpha_+}w_{-,m_2}^{\frac{\alpha_-}{2}}w_{-,m_3}^{\frac{\alpha_-}{2}}w_{-,m_4}^{\frac{1}{2}},\nonumber\\
b_{\underline{m}}^{(2)}&=&w_{-,m_1}^{\alpha_-}w_{-,m_2}^{\frac{\alpha_+}{2} +\frac{1}{4}}w_{-,m_3}^{\frac{\alpha_+}{2} -\frac{1}{4}}w_{-,m_4}^{\frac{1}{2}},\nonumber\\
b_{\underline{m}}^{(3)}&=&w_{-,m_1}^{\alpha_1+1}w_{-Sgn ( \alpha_1),m_2}^{|\alpha_1|}w_{-Sgn(\alpha_2),m_3}^{|\alpha_2|}w_{-,m_4}^{\frac{1}{2}}, \nonumber\\
b_{\underline{m}}^{(4)}&=&w_{-,m_1}^{\alpha_3+\frac{1}{2}}w_{-Sgn(\alpha_3),m_2}^{|\alpha_3|}w_{-,m_3}^{2\alpha_1+1}w_{-,m_4}^{\frac{1}{2}},\nonumber\\
b_{\underline{m}}^{(5)}&=&w_{-,m_1}^{\frac{1}{4g}+\frac{9g}{4}}w_{-Sgn ( \alpha_4),m_2}^{|\alpha_4|}w_{-Sgn(\alpha_4),m_3}^{|\alpha_4|}w_{-,m_4}^{\frac{1}{2}} \nonumber\\
b_{\underline{m}}^{(6)}&=&w_{-Sgn(\alpha_5),m_1}^{|\alpha_5|}w_{-,m_2}^{\alpha_6-\frac{3}{4}}w_{-,m_3}^{\alpha_6+\frac{3}{4}}w_{-,m_4}^{\frac{1}{2}},\nonumber
\end{eqnarray}
with
\begin{eqnarray}
\alpha_{\pm}&=&\frac{1}{4g}\pm \frac{g}{4},\nonumber\\
\alpha_1&=&\frac{1}{8g}+\frac{3g}{8}-\frac{1}{2},\,\,\alpha_2=\frac{1}{8g}-\frac{3g}{8},\nonumber\\
\alpha_3&=&\frac{1}{8g}-\frac{3g}{8}-\frac{1}{4},\,\, \alpha_{4}=\frac{1}{8g}-\frac{9g}{8},\nonumber\\
\alpha_5&=&\frac{1}{4g}-\frac{9g}{4},\qquad \alpha_6=\frac{1}{8g}+\frac{9g}{8}.\nonumber
\end{eqnarray}


\begin{thebibliography}{99}
\bibitem{wigner}Wigner E 1934 {\it Phys. Rev.} {\bf 46} 1002
\bibitem{vignale} Giuliani G F and Vignale G 2005 {\it Quantum Theory
 of the Electron Liquid} (Cambridge University Press, Cambridge).
\bibitem{wigmol1} Reimann S M and Manninen M 2002 {\it Rev. Mod. Phys.} {\bf
 74} 1283
\bibitem{wigmol2} Yannouleas C and Landman U 2007 {\it Rep. Prog. Phys.} {\bf 70} 2067
\bibitem{koudots} Kouwenhoven L P, Marcus C M, McEuen P L,
 Tarucha S, Westervelt R M, and Wingreen N S 1997 in {\em Electron
  transport in quantum dots}, NATO Advanced Studies Institute,
 Series E: Applied Science, edited by L. L. Sohn, L. P. Kouwenhoven,
 and G. Sch\"on (Kluwer, Dordrecht) 105
\bibitem{2Dnum2} Hawrylak P and Pfannkuche D 1993 {\it Phys. Rev. Lett.}
 {\bf 70} 485
\bibitem{Egger99} Egger R, Hausler W, Mak C H, and Grabert H 1999 {\it Phys. Rev. Lett.}
 {\bf 82} 3320
\bibitem{2Dnum3} Tavernier M B, Anisimovas E, Peeters F M,
 Szafran B, Adamowski J, and Bednarek S 2003 {\it Phys. Rev. B} {\bf 68} 205305
\bibitem{2Dnum5} Rontani M, Cavazzoni C, Bellucci D, and
 Goldoni G 2006 {\it J. Chem. Phys.} {\bf 124} 124102
\bibitem{2Dnum6} Harju A, Saarikoski H, and R\"as\"anen E,
 2006 {\it Phys. Rev. Lett.} {\bf 96} 126805
\bibitem{2Dnum7} Yannouleas C and Landman U 2000 {\it Phys. Rev. Lett.} {\bf
 85} 1726
\bibitem{serra} Puente A, Serra L, and Nazmitdinov R G,
 {\it Phys. Rev. B} {\bf 69} 125315
\bibitem{2Dnum8} De Giovannini U, Cavaliere F, Cenni R,
 Sassetti M, and Kramer B 2008 {\it Phys. Rev. B} {\bf 77} 035325
\bibitem{2Dnum10} Cavaliere F, De Giovannini U, Sassetti M and
 Kramer B 2009 {\it New J. Phys.} {\bf 11} 123004
\bibitem{maxwf} Rontani M, Molinari E, Maruccio G, Janson M,
 Schramm A, Meyer C, Matsui T, Heyn C, Hansen W, and
 Wiesendanger R 2007 {\it J. Appl. Phys.} {\bf 101} 081714
\bibitem{giamarchi}Giamarchi T 2004 \textit{Quantum Physics in One
 Dimension} (Oxford Science Publications, Oxford)
\bibitem{schulz}Schulz H J 1993 {\it Phys. Rev. Lett.} {\bf 71} 1864
\bibitem{kramer}H\"ausler and B. Kramer 1993 {\it Phys. Rev. B} \textbf{47} 16353
\bibitem{xia}Gao X 2012 {\it Phys. Rev. A} \textbf{86} 023616
\bibitem{bortz}S\"offing S A, Bortz M, Schneider I, Struck A,
 Fleischhauer F, and Eggert S, {\it Phys. Rev. B} \textbf{79} 195114
\bibitem{1Dwig}Meyer J S and Matveev K A 2009 {\it J. Phys:
 Condens. Matter} {\bf 21} 023203
\bibitem{mantelli}Mantelli D, Cavaliere F, and Sassetti M 2012 {\it J. Phys.: Condens. Matter} {\bf24} 432202
\bibitem{safi}Safi I, and Schulz H J 1999 {\it Phys. Rev. B} \textbf{59} 3040
\bibitem{sablikov}Gindikin Y and Sablikov V A 2007 {\it Phys. Rev. B}
 \textbf{76} 045122
\bibitem{pederiva} Agosti D, Pederiva F, Lipparini E, and Takayanagi K 1998 {\it Phys. Rev. B} {\bf 57} 14869 
\bibitem{bedu} Bed\"urftig G, Brendel B, Frahm H, and Noack R M 1998 {\it Phys. Rev. B} {\bf 58} 10225
\bibitem{szafran} Szafran B, Peeters F M, Bednarek S, Chwiej T,
 and Adamowski J 2004 {\it Phys. Rev. B} {\bf 70} 035401
\bibitem{wire3}Mueller E J 2005 {\it Phys. Rev. B} \textbf{72} 075322
\bibitem{polini} Abedinpour S H, Polini M, Xianlong G, and
 Tosi M P 2007 {\it Phys. Rev. A} {\bf 75} 015602
\bibitem{shulenburger} Shulenburger L, Casula M, Senatore G, and
 Martin R M 2008 {\it Phys. Rev. B} {\bf 78} 165303
\bibitem{secchi1}Secchi A and Rontani M 2009 {\it Phys. Rev. B} \textbf{80} 041404(R)
\bibitem{sgm2}Qian J, Halperin B I, and Heller E J 2010 {\it Phys. Rev. B} \textbf{81} 125323
\bibitem{astrak}Astrakharchik G E and Girardeau M D 2011 {\it Phys. Rev. B} {\bf 83} 153303
\bibitem{burke}Magyar R J and Burke K 2004 {\it Phys. Rev. A} {\bf70} 032508
\bibitem{polini2}Xianlong G, Polini M, Asgari R, and Tosi M P 2006 {\it Phys. Rev. A} {\bf73} 033609
\bibitem{silva}Lima N A, Silva M F, Oliveira L N, and Capelle K 2003 {\it Phys. Rev. Lett.} {\bf90} 146402
\bibitem{AFM}Traverso Ziani N, Cavaliere F, and Sassetti M 2012 {\it Phys. Rev. B} \textbf{86} 125451
\bibitem{sgm1}Zhang L M and Fogler M M 2006 Nano Lett. \textbf{6} 2206
 \bibitem{linear}Boyd E E and Westervelt R M, {\it Phys. Rev. B} \textbf{84} 205308
\bibitem{lee}Lee J, Eggert S, Kim H, Kahng S -J, Shinohara H, and Kuk Y, {\it Phys. Rev. Lett.} \textbf{93} 166403
\bibitem{glazcap} Ussishkin I and Glazman L I 2004 {\it Phys. Rev. Lett} {\bf 93} 196403
\bibitem{sassetti98} Sassetti M and Kramer B 1998 {\it Phys. Rev. Lett.} \textbf{80} 1485
\bibitem{noi} Traverso Ziani N, Piovano G, Cavaliere F, and Sassetti M 2011 {\it Phys. Rev. B} \textbf{84} 155423
\bibitem{eggertstm} Eggert S 2000 {\it Phys. Rev. Lett.} {\bf 84} 4413
\bibitem{martin} Crepieux A, Guyon R, Devillard P, and Martin T 2003 {\it Phys. Rev. B} \textbf{67} 205408
\bibitem{dolcini}Pugnetti S, Dolcini F, Bercioux D, and Grabert H, {\it Phys. Rev. B} \textbf{79} 035121
\bibitem{nocera} Nocera A, Perroni C A, Marigliano Ramaglia V, and Cataudella V 2012 {\it Phys. Rev. B} {\bf 86} 035420
\bibitem{secchi2}Secchi A and Rontani M 2012 {\it Phys. Rev. B} \textbf{85} 121410
\bibitem{haldanefluid}Haldane F D M 1981 {\it Phys. Rev. Lett.} \textbf{47} 1840
\bibitem{voit}Voit J 1995 {\it Rep. Prog. Phys.} \textbf{58} 977
\bibitem{delft}Delft J v and Schoeller H 1998 {\it Annalen Phys.} \textbf{7} 225
\bibitem{eggerimp} Egger R and Grabert H 1997 {\it Phys. Rev. Lett.} {\bf 79} 3463
\bibitem{braggioepl} Braggio A, Grifoni M, Sassetti M, and Napoli F 2000 {\it Europhys. Lett.} {\bf 50} 236
\bibitem{iotobias} Kleimann T, Cavaliere F, Sassetti M, and Kramer B 2002 {\it Phys. Rev. B} {\bf 66} 165311
\bibitem{kim} Kim J U, Krive I V, and Kinaret J M 2003 {\it Phys. Rev. Lett.} {\bf 90} 176401
\bibitem{ioale1} Cavaliere F, Braggio A, Stockburger J T, Sassetti M, and Kramer B 2004 {\it Phys. Rev. Lett.} {\bf 93} 036803
\bibitem{milena} Mayrhofer L and Grifoni M 2007, {\it Eur. Phys. J. B} \textbf{56} 107; Cuniberti G, Sassetti M, and Kramer B 1997 {\it Phys. Rev. B} {\bf57} 1515
\bibitem{k05}Schulz H J 1990 {\it Phys. Rev. Lett.} {\bf 64} 2831
\bibitem{fiete1}Fiete G A, Le Hur K, and Balents L 2006 {\it Phys. Rev. B} \textbf{73} 165104
\bibitem{075}Matveev K A 2004 {\it Phys. Rev. Lett.} {\bf 92} 106801
\bibitem{pscr}Traverso Ziani N, Cavaliere F, Piovano G and Sassetti M 2012 {\it Phys. Scr.} {\bf T151} 014041
\bibitem{fabrizio} Fabrizio M and Gogolin A O 1995 {\it Phys. Rev. B} \textbf{51} 17827
\bibitem{flensberg}H. Bruus, and K. Flensberg, \textit{Many-Body Quantum
Theory in Condensed Matter Physics: An Introduction},
Oxford University Press (2004).
\bibitem{bercioux} Bercioux D, Buchs G, Grabert H, and Gr\"oning O, {\it Phys. Rev. B} \textbf{83} 165439
\bibitem{tunneling} Prange R E 1963 {\it Phys. Rev.} \textbf{131} 1083
\bibitem{blum} Blum K 1981 \textit{Density Matrix Theory and Applications},
 Plenum Press; Grifoni M, Sassetti M, and Weiss U 1996 {\it Phys. Rev. E} \textbf{53} R2033; Paladino E, Sassetti M, Falci G, and Weiss U 2008 {\it Phys. Rev. B} \textbf{77} 041303
\bibitem{nonlin1}Haupt F, Cavaliere F, Fazio R, and Sassetti M 2006 {\it Phys. Rev. B} \textbf{74} 205328; Merlo M, Haupt F, Cavaliere F, and Sassetti M 2008 {\it New J. Phys.} \textbf{10} 023008; Cavaliere F, Mariani E, Leturcq R, Stampfer C, and Sassetti M 2010 {\it Phys. Rev. B} \textbf{81} 201303(R)
\bibitem{master}Furusaki A 1998 {\it Phys. Rev. B} \textbf{57} 7141
\bibitem{nonlin2}Piovano G, Cavaliere F, Paladino E, and Sassetti M 2011 {\it Phys. Rev. B} \textbf{83} 245311
\bibitem{STMWIRE} Eder C, Smoliner J, B\"ohm G and Weimann G 1996 {\it Semicond. Sci. Technol.} \textbf{11} L1239
\bibitem{STMCNT} Odom T W, Huang J-L, and Lieber C M 2002 {\it J. Phys.: Condens. Matter} {\bf 14} R145
\end{thebibliography}
\end{document}